\documentclass[10pt,aps,prc,twocolumn,groupedaddress,floatfix,nofootinbib]{revtex4-1}% KS
\usepackage{times}
\usepackage[T1]{fontenc}

\usepackage{multirow}

\usepackage{amsmath,amssymb,mathrsfs}
\usepackage{bm}% bold math
\usepackage{ascmac}
\usepackage{geometry}
\usepackage{graphicx}
\begin{document}

\title{
%Exploring orientation effects in multinucleon transfer processes with deformed target:\\
%the $^{16}$O+$^{154}$Sm reaction at around and above the Coulomb barrier
Reaction mechanism study for multinucleon transfer processes in collisions
of spherical and deformed nuclei at energies near and above the Coulomb barrier:
The $^{16}$O+$^{154}$Sm reaction}
% Force line breaks with \\

\author{B.J.~Roy$^{1,2}$}\email{bjroy@barc.gov.in, bidyutr2003@gmail.com}
\author{S.~Santra$^{1,2}$}
\author{A.~Pal$^1$}
\author{H.~Kumawat$^1$}
\author{S.~K.~Pandit$^1$}
\author{V.~V.~Parkar$^1$}
\author{K. Ramachandran$^1$}
\author{K.~Mahata$^1$}

\affiliation{$^1$Nuclear Physics Division, Bhabha Atomic Research Centre, Mumbai 400 085, India}
\affiliation{$^2$Homi Bhabha National Institute, Anushakti Nagar, Mumbai 400 094, India}

\author{K.~Sekizawa$^{3,4}$}\email{sekizawa@phys.titech.ac.jp}
\affiliation{$^3$Department of Physics, School of Science, Tokyo Institute of Technology, Tokyo 152-8551, Japan}
\affiliation{$^4$Nuclear Physics Division, Center for Computational Sciences, University of Tsukuba, Ibaraki 305-8577, Japan}

\date{April 5, 2022}

\begin{abstract}
\begin{description}
\item[Background]% This part should describe the context needed to understand what the paper is about.
Multinucleon transfer reactions at energies around the Coulomb barrier offer a vital opportunity
to study rich physics of nuclear structure and dynamics, e.g., single-particle level structure
and quantum shells, mass/charge equilibration processes, energy dissipation, as well as secondary
decays via particle emission or fission. Despite the continuous development in the field, we have
still limited knowledge about how deformation---one of the representative nuclear structures---affects
multinucleon transfer reactions.

\item[Purpose]% This part should state the purpose of the present paper.
To develop our understanding of the reaction mechanism and to
shed light on the effect of deformation in multinucleon transfer processes,
we study the $^{16}$O+$^{154}$Sm reaction at $E_{\rm lab}$\,=\,85\,MeV
(near the Coulomb barrier) and 134\,MeV (substantially above the
Coulomb barrier), where the target nucleus, $^{154}$Sm, is a well-established,
deformed nucleus.

\item[Methods]% This part describes the methods used in the paper.
We have performed experiments on the $^{16}$O+$^{154}$Sm reaction at the BARC-TIFR
pelletron-Linac accelerator facility, Mumbai, India, measuring angular distributions
and $Q$ value spectra for various transfer products. The measured cross sections
have been analyzed along with theoretical calculations based on the time-dependent
Hartree-Fock (TDHF) theory, together with a statistical model for secondary deexcitation
processes, GEMINI++.

\item[Results]% This part should summarize the results of the methods described in the previous tag.
Angular distributions for elastic scattering and for various transfer channels were measured over a wide angular range.
The $Q$-value- and angle-integrated isotope production cross sections have been extracted
from the measured angular distributions. We have successfully obtained production cross sections
for various isotopes for $E_{\rm lab}$\,=\,85\,MeV, while only for four isotopes could be deduced
for $E_{\rm lab}$\,=\,134\,MeV due to present experimental limitations. For the lower incident energy
case, we find a reasonable agreement between the measurements and the TDHF calculations for
a-few-nucleon transfer channels; whereas TDHF underestimates cross sections for many-nucleon
transfers, consistent with earlier works. On the other side, we find that calculated cross sections
for secondary reaction products for the higher incident energy case, qualitatively explains the measured
trends of isotopic distributions observed for the lower energy. The latter observation indicates possible
underestimation of excitation energies in the present TDHF+GEMINI analysis. Although certain orientation
effects were observed in TDHF results, it turns out to be difficult to disentangle them from
the $Q$-value- and angle-integrated production cross sections.

\item[Conclusions]% This part should state the conclusions of the paper.
The present analysis highlights the deep-inelastic character of multinucleon transfer processes
and importance of secondary deexcitation processes. We showed that the orientation effect in multinucleon
transfer processes in the $^{16}$O+$^{154}$Sm reaction is rather weak and hard to disentangle from
the present measured data. Further systematic investigations especially in the sub-barrier energy regime,
where the data would be more sensitive to single-particle properties, would be required to uncover effects
of nuclear deformation on multinucleon transfer processes in low-energy heavy-ion reactions.
\end{description}
\end{abstract}

%\pacs{25.70.Hi, 24.10.-i, 25.70.Bc, 21.60.Jz}% PACS, the Physics and Astronomy Classification Scheme.
%\keywords{}
% KS: PACS numbers are no longer necessary for PRC

\maketitle

\section{INTRODUCTION}\label{sec:intro}

The deformation of atomic nuclei plays a significant role in low-energy heavy-ion reactions.
A well-established, intuitive example is the orientation dependence of the Coulomb barrier
height, which affects dramatically the fusion cross section (see, e.g., Ref.~\cite{Hagino(2012)}).
In quasifission processes, characterized by a massive nucleon transfer with zeptosecond contact
time without the compound nucleus formation, the orientation of a deformed nucleus substantially
alters the reaction dynamics \cite{Wakhle(2014),Oberacker(2014),Umar(2015),KS_KY_Ni-U,
Umar(2016),Zheng(2018),Guo(2018),Godbey(2019)}. When a collision occurs at equatorial
side of a prolately deformed actinide nucleus, for instance, a compact system is formed in the
course of collision, leading to longer contact times with larger amount of nucleon transfer toward
the mass equilibrium of the system. In tip collisions, on the other hand, an elongated dinuclear
system is formed, leading to shorter contact times, while stronger shell effects are often observed.
Besides, orientation dependent inverse (or anti-symmetrizing) quasifission processes have been
reported for $^{232}$Th+$^{250}$Cf \cite{Kedziora(2010)} and $^{238}$U+$^{124}$Sn
\cite{KS_U-Sn}. As compared to the orientation effects in damped collisions of heavy ions,
like quasifission processes, those in multinucleon transfer (MNT) processes in peripheral
collisions have been less investigated so far.

There has been increasing interest in recent years in the study of MNT processes in
heavy-ion reactions at energies around and above the Coulomb barrier (see, e.g., review
papers \cite{Corradi(2009),Zhang(2018)review,Sekizawa(2019),Adamian(2020)} and references
therein). Partly because, the MNT reaction plays a crucial role for understanding nucleon-nucleon
correlations and for giving an opportunity to access a wide variety of nuclear structures in the far-off
stability region. Besides, the MNT reaction between heavy ions is expected to be an efficient
approach for production of neutron-rich heavy nuclei whose production is difficult by other methods
\cite{Zagrebaev(2008),Zagrebaev(2011),Zagrebaev(2013),Karpov(2017),Saiko(2019),
Zhu(2015),Zhu(2016),Zhu(2017)1,Zhu(2017)2,Feng(2017),Zhang(2018),Chen(2020),
Zhao(2015),Wang(2016),Zhao(2016),Li(2019),Zhao(2021)}, supported with promising experimental
evidence \cite{Watanabe(136Xe+198Pt)}. The production of neutron-rich superheavy nuclei
in the predicted island of stability ($Z$\,=\,114, 120, or 126, $N$\,=\,184) is highly desired \cite{
Hofmann(2000),Oganessian(2017)}, because, in addition to fundamental interest in nuclear structures
such as shell evolution \cite{Otsuka(2018)} and shape transitions \cite{Cejnar(2010),Heyde(2011)},
it would provide a new stringent constraint for microscopic theories. The investigation
of MNT processes is an important project at current and future RIB facilities such as
RIBF (RIKEN, Japan) \cite{Sakurai(2018)}, HIRFL-CSR and HIAF (IMP, China) \cite{HIAF},
RAON (RISP, Korea) \cite{RAON}, DRIB (FLNR, Russia), SPIRAL2 (GANIL, France) \cite{Gales(2007)},
FAIR (GSI, Germany) \cite{FAIR}, SPES (INFN, Italy) \cite{SPES}, and FRIB (MSU, USA) \cite{FRIB},
and so forth. It is important to provide a reliable prediction of the optimum reaction condition,
such as projectile-target combinations and collision energies, to guide experiments to
\textit{terra incognita}. Since the majority of nuclei are actually deformed in their
ground state, it is naturally of crucial importance to explore possible effects of mutual
orientations on transfer dynamics that could also be an important factor to optimize.

In our preceding studies, we have investigated the MNT mechanism in $^{18}$O+$^{206}$Pb
\cite{Bidyut(2015)} and $^{16}$O+$^{27}$Al \cite{Bidyut(2018)} reactions at
energies above the Coulomb barrier. We carried out analyses of the experimental data
along with theoretical calculations based on the microscopic framework of the time-dependent
Hartree-Fock (TDHF) theory. As the theory based on the independent-particle picture,
a detailed comparison with experimental data can offer useful information on multi-particle
correlations as well as deformation/orientation dependence of MNT dynamics. In Refs.~\cite{Bidyut(2015),Bidyut(2018)},
we found reasonable agreements between measurements and TDHF calculations for
$Q$-value- and angle-integrated transfer cross sections for both systems, where significance
of secondary particle emissions were underlined. In those works, $^{18}$O and $^{27}$Al nuclei
were found to be deformed in prolate and oblate shapes, respectively, in their Hartree-Fock ground
state (without pairing correlations). To explore possible orientation effects on MNT processes,
TDHF calculations were performed for different initial orientations of those deformed nuclei,
and signatures of orientation dependence were observed: \textit{i}) In the $^{18}$O+$^{206}$Pb
reaction \cite{Bidyut(2015)}, we found that neutrons could be transferred towards the
opposite direction to the one expected from the charge asymmetry of the system, when
deformation axis is set initially aligned to the impact parameter vector; \textit{ii}) In the
$^{16}$O+$^{27}$Al reaction \cite{Bidyut(2018)}, we found that the colliding system
shows a strong tendency towards mass equilibration, when deformation axis is set
perpendicular to the reaction plane. Although those observations are intriguing, it was
not possible to see the effects in the experimental data. Also, it might be an artifact of
the neglected pairing, which causes unrealistic deformation in the Hartree-Fock ground states.

A systematic investigation of MNT reactions with experimental data using different
projectile-target combinations would be necessary for a better understanding of the
deformation effect in the reaction mechanism. With such a motivation in mind,
here we report investigation of MNT processes in the $^{16}$O+$^{154}$Sm
reaction, where the target nucleus, $^{154}$Sm, is a well-deformed nucleus
($\beta_2\simeq 0.34$, $E_{4+}/E_{2+}\simeq3.25$ \cite{NNDC}). The
experiment was carried out at the heavy-ion accelerator Pelletron-Linac facility, Mumbai, India.
We set bombarding energies as $E_{\rm lab}$\,=\,85\,MeV (near the Coulomb barrier, 1.1$V_{\rm B}$)
and $E_{\rm lab}$\,=\,134\,MeV (substantially above the barrier, 1.7$V_{\rm B}$).
TDHF calculations at these two energies were performed and detailed comparison
with the measurements has been made for understanding of MNT dynamics.

The article is organized as follows.
In Sec.~\ref{sec:expt}, we describe details of the measurement and present our experimental data.
In Sec.~\ref{sec:TDHF}, the results of theoretical (TDHF) calculations are presented and compared with the experimental data.
A summary is given in Sec.~\ref{sec:summary}.

\section{EXPERIMENTAL DETAILS}\label{sec:expt}

%&&&&&&&&&&&&&&&&&&&&&&&&&&&&&&&&&&&&&&&&&&&&&&&&&&
\begin{figure}[t]
\begin{center}
\includegraphics[width=\columnwidth]{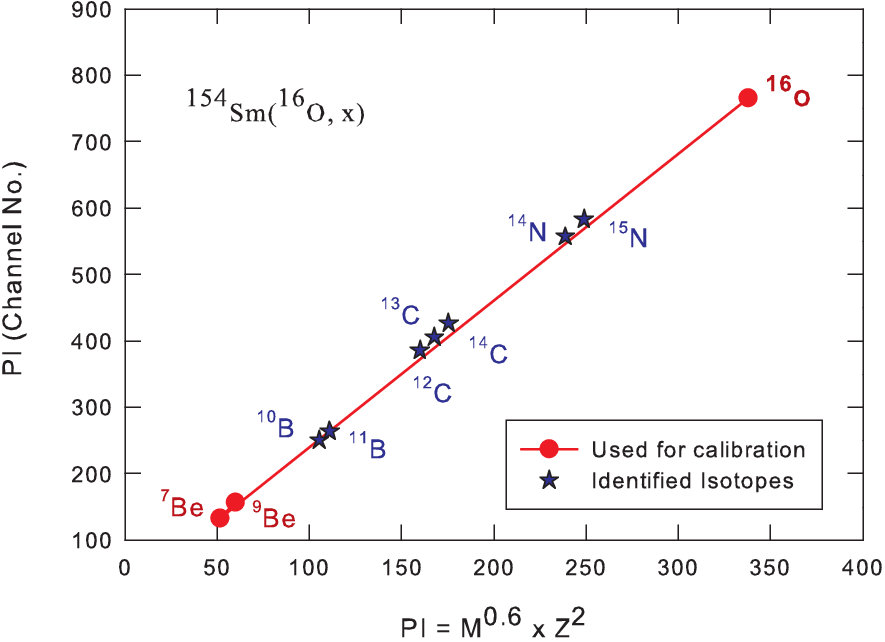}
\caption{
Typical particle identification (PI) spectrum obtained from the
$^{16}$O+$^{154}$Sm reaction at $E_{\rm lab}$\,=\,85\,MeV
for various projectile-like fragments.
}\vspace{-3mm}
\label{PISm}
\end{center}
\end{figure}
%&&&&&&&&&&&&&&&&&&&&&&&&&&&&&&&&&&&&&&&&&&&&&&&&&&

Experiments were performed at the BARC-TIFR pelletron-Linac accelerator facility,
Mumbai, with $^{16}$O beam. Target used was enriched $^{154}$Sm
foil of thickness 240\,$\mu$g/cm$^2$. The $^{16}$O+$^{154}$Sm
reaction was studied at $E_{\rm lab}$\,=\,85 and 134\,MeV. Energy
uncertainty of LINAC beam was $\pm0.5$\,MeV. Samarium target was prepared at Radio Chemistry Division, Bhabha
Atomic Research Centre (BARC), Mumbai, by electro-deposition method
on $^{27}$Al backing with backing-foil thickness of 540\,$\mu$g/cm$^2$.
Measurements with the samarium target at all the angles were repeated by
replacing samarium with pure aluminium target that was used as backing
in order to subtract any contribution in the angular distribution data that
may be contributed from aluminium backing. For detection and identification
of reaction products [projectile-like fragments (PLFs)] several silicon surface
barrier detector telescopes of appropriate thickness in $\Delta E$-$E$
configuration were used. The experimental setup and measurement
details are very similar to the ones that we used in our earlier works
\cite{Bidyut(2015),Bidyut(2018)}, and details can be found in these
references. A good charge and mass separation was achieved in $\Delta E$-$E$
spectrum and PLFs were identified, following standard particle identification
(PI) technique as discussed in \cite{Bidyut(2015),Bidyut(2018)} (see Fig.~\ref{PISm}).

%&&&&&&&&&&&&&&&&&&&&&&&&&&&&&&&&&&&&&&&&&&&&&&&&&&
\begin{figure}[t]
\begin{center}
\includegraphics[width=0.94\columnwidth]{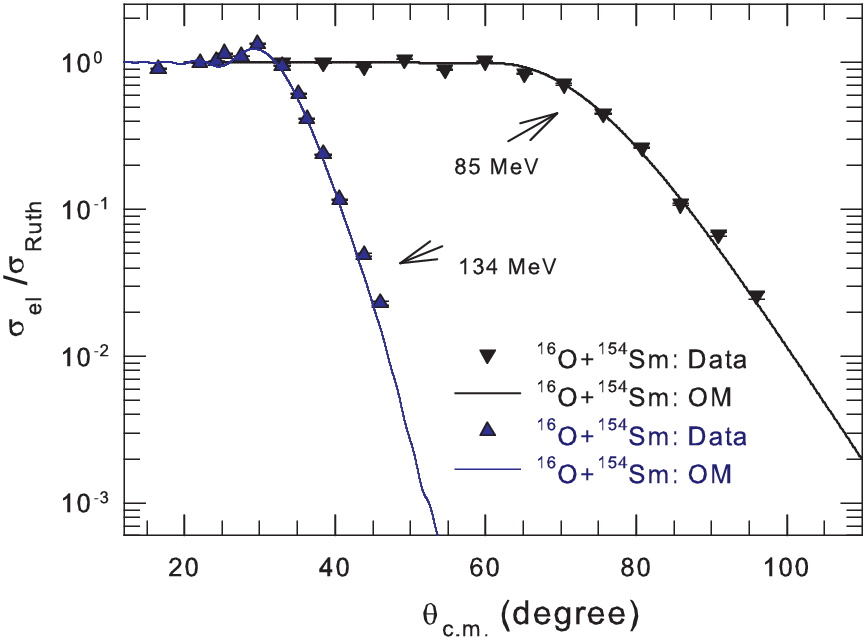}\vspace{-3mm}
\caption{
The ratio of elastic scattering to the Rutherford cross sections for the
$^{16}$O+$^{154}$Sm reaction at $E_{\rm lab}$\,=\,85 and 134\,MeV
(shown by downward and upward triangles, respectively) as a function
of the scattering angle in the center-of-mass frame in degrees. Solid lines
represent results of optical-model (OM) calculations with the \texttt{SFRESCO}
code \cite{SFRESCO} for these two energies. The potential parameters
obtained from the best fitting to the data are given in Table~\ref{TableOM}.
}\vspace{-3mm}
\label{elast1}
\end{center}
\end{figure}
%&&&&&&&&&&&&&&&&&&&&&&&&&&&&&&&&&&&&&&&&&&&&&&&&&&

%//////////////////////////////////////////////////////////////////////////
\begin{table}[b]
\caption{
Potential parameters for the $^{16}$O+$^{154}$Sm reaction at
$E_{\rm lab}$\,=\,85 and 134\,MeV, obtained from the optical-model analysis
of the measured elastic-scattering angular distribution with the \texttt{SFRESCO}
code \cite{SFRESCO}. $V$, $r$, and $a$ ($V_i$, $r_i$, and $a_i$) represent depth,
radius parameter, and diffuseness of the real (imaginary) part of the potential, respectively.
The cumulative reaction cross section, $\sigma_{\rm R}$, is also shown at the bottom row.
}
\begin{center}
\begin{tabular*}{\columnwidth}{@{\extracolsep{\fill}}ccc}
\hline\hline\\[-2.8mm]
Potential  & $^{16}$O+$^{154}$Sm & $^{16}$O+$^{154}$Sm \\
parameters & 85 MeV  & 134 MeV\\
\hline\\[-2.5mm]
$V$ (MeV) & 24.0   & 24.1\\[0.2mm]
$r$ (fm)  & 1.24 & 1.24\\[0.2mm]
$a$ (fm)  & 0.59  & 0.73 \\[0.2mm]
$V_i$ (MeV) & 17.6    &  16.5\\[0.2mm]
$r_i$ (fm)  & 1.26  & 1.25 \\[0.2mm]
$a_i$ (fm)  & 0.68 & 0.72 \\
\hline\\[-3mm]
$\sigma_{\rm R}$ (mb)  & 1182 & 2357 \\
\hline\hline
\end{tabular*}\vspace{-3mm}
\label{TableOM}
\end{center}
\end{table}
%//////////////////////////////////////////////////////////////////////////

Elastic-scattering angular distributions were measured and optical-model
calculations have been carried out. Results are shown in Fig.~\ref{elast1}
and Table~\ref{TableOM}. In the figure, data are plotted along with statistical
errors and in most of the cases the error bars are within the data symbols.
In order to extract potential parameters listed in Table~\ref{TableOM},
the optical-model search program \texttt{SFRESCO} \cite{SFRESCO} has
been used and as usual a volume Woods-Saxon form is adapted for the
real and imaginary parts of the potential.

At first, we carried out the optical-model analysis with the code \texttt{SFRESCO}
for the $E_{\rm lab}$\,=\,85\,MeV case. The potential parameters that gave best
fit (Chi-Square minimization fit) to the presently measured elastic-scattering angular
distribution at $E_{\rm lab}$\,=\,85\,MeV are shown in the second column of
Table~\ref{TableOM}. The cumulative reaction cross section is also listed in the
bottom row of Table~\ref{TableOM}. These parameters were then used as a
starting potential for analyzing the $E_{\rm lab}$\,=\,134\,MeV data. Some
variation of the potential parameters were needed to get a reasonably good
agreement with the measured angular distribution at 134\,MeV (Fig.~\ref{elast1}).
The best fit potential parameters for $E_{\rm lab}$\,=\,134\,MeV are listed
in column~3 of Table~\ref{TableOM}. Significant increase in the cumulative
reaction cross section, as expected, is observed with increasing the incident
energy from 85 to 134\,MeV (see the bottom row of Table~\ref{TableOM}).

%&&&&&&&&&&&&&&&&&&&&&&&&&&&&&&&&&&&&&&&&&&&&&&&&&&
\begin{figure}[t]
\begin{center}
\includegraphics[width=\columnwidth]{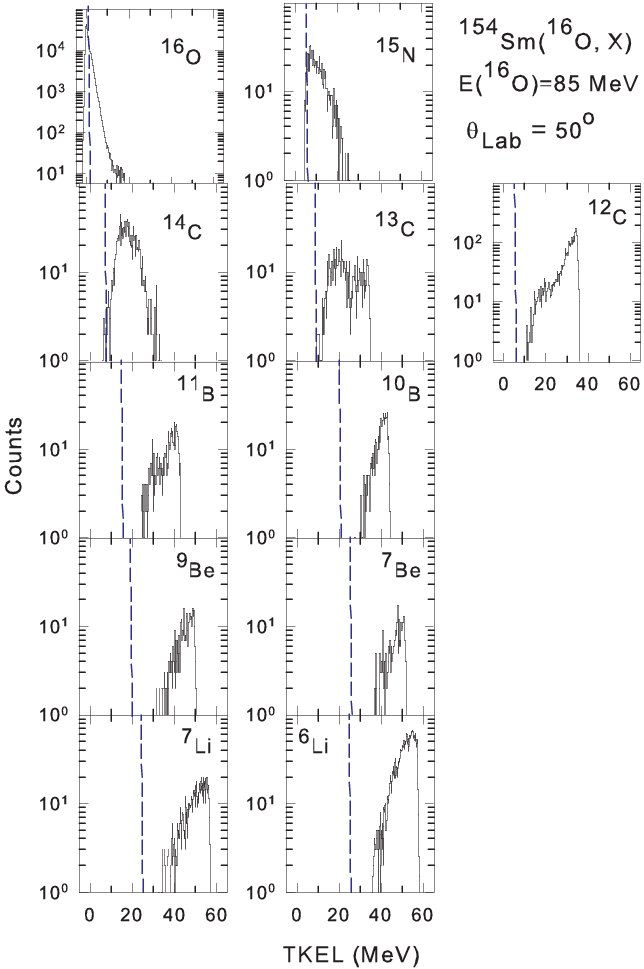}\vspace{-3mm}
\caption{
Experimental total kinetic energy loss (TKEL) distributions (histogram plot)
for various projectile-like fragments in the $^{16}$O+$^{154}$Sm reaction
at $E_{\rm lab}$\,=\,85\,MeV, at a fixed scattering angle of $\theta_{\rm lab}$\,=\,50$^\circ$.
Vertical dashed lines indicate the ground-state $Q$ values.
}\vspace{-3mm}
\label{TKEL85}
\end{center}
\end{figure}
%&&&&&&&&&&&&&&&&&&&&&&&&&&&&&&&&&&&&&&&&&&&&&&&&&&

%&&&&&&&&&&&&&&&&&&&&&&&&&&&&&&&&&&&&&&&&&&&&&&&&&&
\begin{figure}[t]
\begin{center}
\includegraphics[width=\columnwidth]{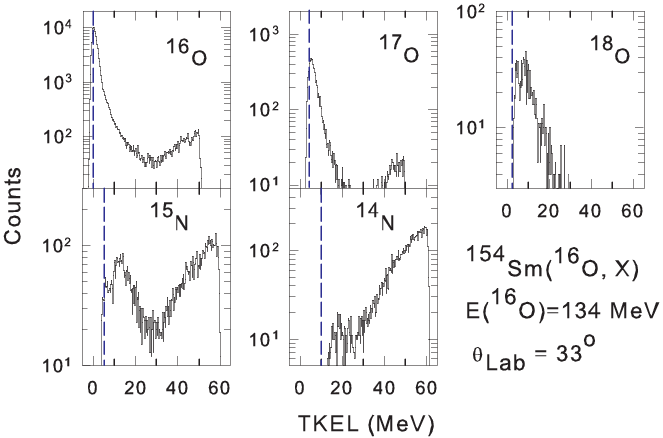}\vspace{-3mm}
\caption{
Same as Fig.~\ref{TKEL85}, but for $E_{\rm lab}$\,=\,134\,MeV
at a fixed scattering angle of $\theta_{\rm lab}$\,=\,33$^\circ$.
}\vspace{-3mm}
\label{TKEL134}
\end{center}
\end{figure}
%&&&&&&&&&&&&&&&&&&&&&&&&&&&&&&&&&&&&&&&&&&&&&&&&&&

We have derived total kinetic energy loss (TKEL) from our measured $Q$-value
spectra, assuming a pure binary process as detailed in Ref.~\cite{Bidyut(2015)}
and references therein. Obtained TKEL distributions for various transfer channels
are shown in Figs.~\ref{TKEL85} and \ref{TKEL134} for the $E_{\rm lab}$\,=\,85-
and 134-MeV cases, respectively. For $E_{\rm lab}$\,=\,85\,MeV (Fig.~\ref{TKEL85}),
TKEL is low for few-nucleon transfer channels indicating dominance of quasielastic transfer.
As the number of transferred nucleons increases, there is a gradual shift of the centroid
of energy-loss spectra towards the larger TKEL and reaches to as large as 50\,MeV
for $^6$Li-production ($-5$p, $-5$n) channel indicating significant contributions
from deep-inelastic processes even at this lower energy case. It is to mention that
the sharp fall- or cut-off seen at higher TKEL for some of the PLFs corresponds to
the low kinetic energy part of the PLFs that gets stopped in the $\Delta E$ detector.
There may be some contamination from aluminium backing in the TKEL spectra,
however, their contribution is expected to be in the higher TKEL part because of the
kinematics. In the case of $E_{\rm lab}$\,=\,134\,MeV (Fig.~\ref{TKEL134}),
we find a 2nd bump at higher TKEL for $^{16}$O channel, which is due to aluminum
backing \cite{Bidyut(2018)}. For neutron pickup reactions ($^{17}$O and $^{18}$O PLFs),
quasielastic transfer reactions predominate and the spectra are clean (any contribution
from aluminum backing is hardly visible except a small bump at higher TKEL in the
$^{17}$O spectrum, which is well separated from samarium events due to kinematics). 
For the channels with stripping of few nucleons (e.g., $^{15}$N and $^{14}$N PLFs),
the first peak at low TKEL corresponds to low reaction $Q$-value, which is then
followed by a strong rise at larger TKEL which might be due to $^{27}$Al($^{16}$O, $x$)
reactions \cite{Bidyut(2018)}. With increasing number of transferred nucleons
further the energy spectra shift towards higher excitation energies and we observed
a strong overlap with events from the aluminium target. It was therefore practically
impossible to separate events from the pure samarium target, and hence no effort
was put to extract TKEL spectra from those MNT channels.  

%&&&&&&&&&&&&&&&&&&&&&&&&&&&&&&&&&&&&&&&&&&&&&&&&&&
\begin{figure}[t]
\begin{center}
\includegraphics[width=\columnwidth]{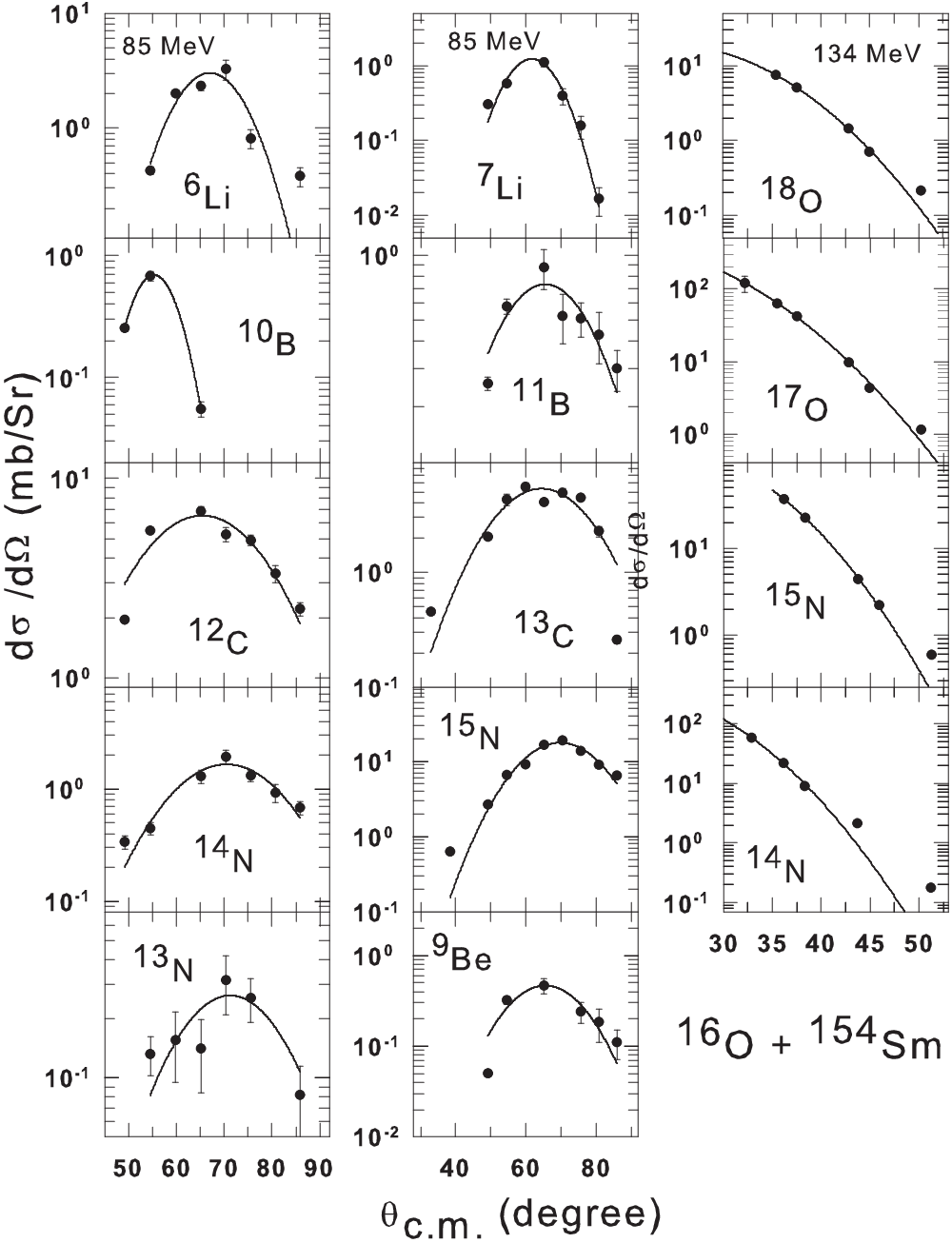}
\caption{
Measured $Q$-value-integrated angular distributions for various transfer
channels in the $^{16}$O+$^{154}$Sm reaction at $E_{\rm lab}$\,=\,85\,MeV
(the first and second columns) and 134\,MeV (the third column). Solid lines
represent 3-parameter Gaussian fits to the data as discussed in the text.
}
\label{AngDistSm}
\end{center}
\end{figure}
%&&&&&&&&&&&&&&&&&&&&&&&&&&&&&&&&&&&&&&&&&&&&&&&&&&

Besides the $Q$-value spectra, we have also measured angular distributions for
various transfer channels. By integrating over the energy, $Q$-value-integrated
differential cross sections for various transfer channels have been obtained, and
the results are shown in Fig.~\ref{AngDistSm}. It is to mention that the pure
neutron-transfer channels in $E_{\rm lab}$\,=\,85-MeV data as well as some
of other transfer channels, though visible at some of the cases at some angles,
were not clearly separated (or having low counts) and hence angular distributions
for those reaction channels could not be extracted. For the $^{16}$O+$^{154}$Sm
reaction at $E_{\rm lab}$\,=\,85~MeV (left and middle columns in Fig.~\ref{AngDistSm}),
the observed angular distributions are in general bell shaped, peaking at an angle
slightly lower than the grazing angle ($\theta_{\rm gr}$\,$\approx$\,80$^\circ$
as obtained from the elastic scattering angular distributions) with a small dependence
on the reaction channel. In the case of $E_{\rm lab}$\,=\,134\,MeV (right column
in Fig.~\ref{AngDistSm}), our measured data are in a limited angular range (also
for a limited number of reaction channels) and the most of the data are beyond
the expected peak angle (see Fig.~\ref{AngDistSm}). For this energy the grazing
angle is $\theta_{\rm gr}$\,$\approx$\,38$^\circ$. Forward angle data were
not clean and contribution from aluminium backing was relatively large, hence no
effort was put to extract cross sections at those angles.

\begin{figure}[t]
\begin{center}
\includegraphics[width=\columnwidth]{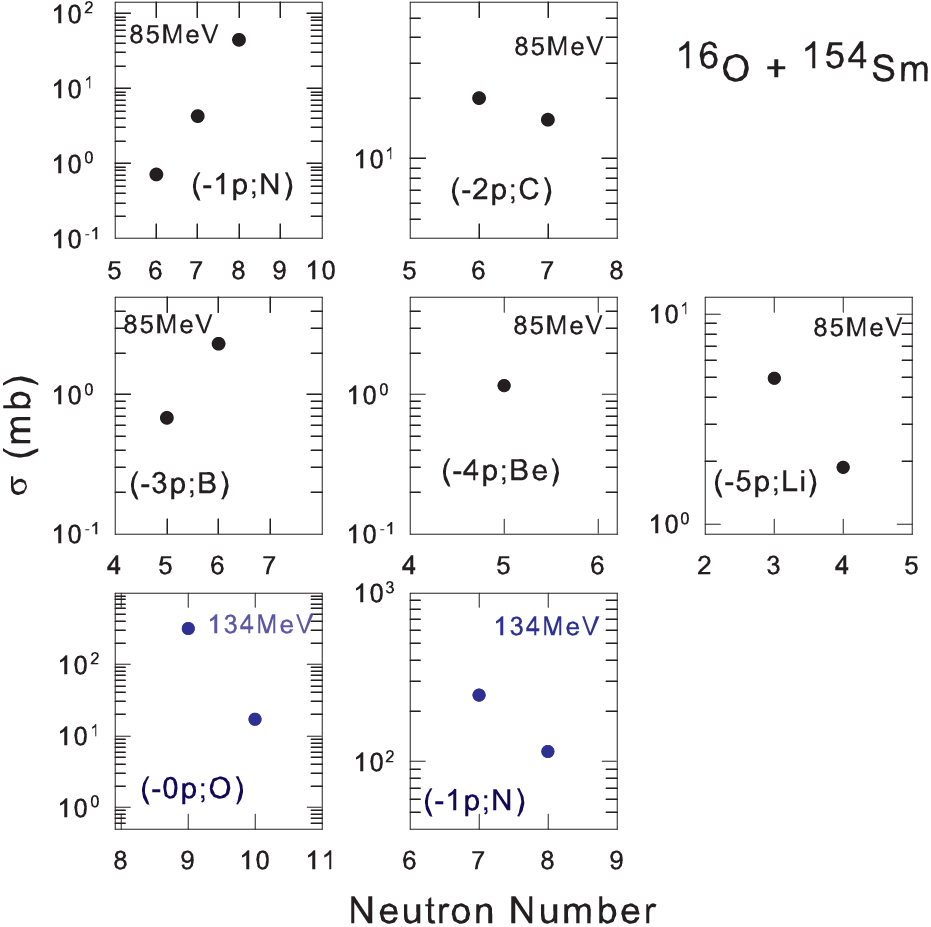}
\caption{
The $Q$-value- and angle-integrated isotope production cross sections for
various proton transfer channels in the $^{16}$O+$^{154}$Sm reaction at
$E_{\rm lab}$\,=\,85\,MeV (the first and second rows) and 134\,MeV (the bottom
row). The change in the number of protons compared with that of the projectile
($Z$\,=\,8) is indicated as ($\pm x$p;~X), where X stands for the corresponding element.
}
\label{SigmaInt154Sm}
\end{center}
\end{figure}

We have then obtained the $Q$-value- and angle-integrated, total production
cross sections for different transfer channels via a 3-parameter Gaussian fit of
the angular distributions, as detailed in Ref.~\cite{Bidyut(2015)}. Obtained isotopic
production cross sections are shown in Fig.~\ref{SigmaInt154Sm} for both
$E_{\rm lab}$\,=\,85\,MeV (top and middle rows) and for 134\,MeV (bottom row).
Although the measured angular distributions are in a limited angular range for
$E_{\rm lab}$\,=\,134\,MeV as shown in Fig.~\ref{AngDistSm}, the 3-parameter
Gaussian fit was applied and total cross sections could be deduced. In the figure,
cross sections for different proton-stripping channels are shown as a function of
the neutron number of PLFs. In the next section, we will discuss the measured
cross sections in comparison with TDHF calculations.

%&&&&&&&&&&&&&&&&&&&&&&&&&&&&&&&&&&&&&&&&&&&&&&&&&&
%&&&&&&&&&&&&&&&&&&&&&&&&&&&&&&&&&&&&&&&&&&&&&&&&&&

\section{TDHF ANALYSIS}\label{sec:TDHF}

To obtain deeper insight into the reaction mechanism and to
explore possible deformation/orientation dependence of transfer
dynamics, TDHF calculations were performed for the $^{16}$O+$^{154}$Sm
reaction at $E_{\rm lab}$\,=\,85 and 134\,MeV. A three-dimensional
parallel TDHF solver was used, which has been continuously developed
since the work by Sekizawa and Yabana \cite{KS_KY_MNT}. The code
has been successfully applied for a number of systems \cite{KS_KY_MNT,
KS_KY_PNP,KS_KY_Ni-U,KS_SH_Kazimierz,KS_GEMINI,KS_U-Sn,KW_KS_DDTDHF},
including various extensions going beyond TDHF \cite{Williams(2018),
KS_KH_TDHF+Langevin,KS_AS_Ni-Pb,AS_KS_Xe-Pb}. In our preceding
studies, we made TDHF analyses of our earlier experimental data for
$^{18}$O+$^{206}$Pb \cite{Bidyut(2015)} and $^{16}$O+$^{27}$Al
\cite{Bidyut(2018)} systems, which serve useful information for the analysis
presented here. The model and calculation details have been described
in the references given above, and here we provide information relevant
to the present analysis. (For details of the theoretical framework and
its applications, see, e.g., review papers \cite{Negele(review),Simenel(review),
Nakatsukasa(2012),Nakatsukasa(2016),TDHF-review(2018),Stevenson(2019),
Sekizawa(2019)} and references therein.)

\begin{figure}[tb]
\begin{center}
\includegraphics[width=60mm]{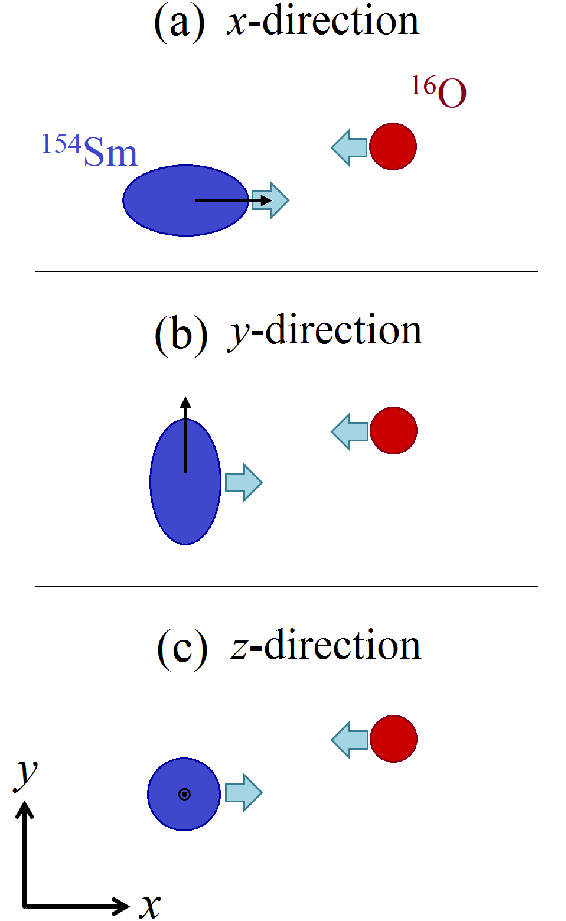}
\end{center}\vspace{-3mm}
\caption{
A schematic illustration of three initial configurations for the $^{16}$O+$^{154}$Sm
reaction used in the TDHF calculations. Red and blue discs represent cross sections
of the density of the projectile and target nuclei, respectively, in the reaction plane
($xy$ plane). Incident direction represented by thick arrows are parallel to the $x$-axis
and the impact parameter vector is parallel to the $y$-axis. The three initial configurations
considered are: (a) symmetry axis set parallel to the collision axis ($x$ axis), (b) symmetry
axis set parallel to the impact parameter vector ($y$ axis), and (c) symmetry axis set
perpendicular to the reaction plane.
}
\label{orientation}
\end{figure}

For the energy density functional, the SLy5 parameter set \cite{Chabanat(1998)}
was used, which is the same as in our preceding studies \cite{Bidyut(2015),Bidyut(2018)}.
The Hartree-Fock ground state of $^{16}$O is of spherical shape, while that of
$^{154}$Sm is largely deformed in prolate shape ($\beta_2\simeq0.32$), exhibiting
a small octupole deformation. To explore orientation dependence of the reaction
mechanism, TDHF calculations were performed for three initial orientations of $^{154}$Sm
as depicted in Fig.~\ref{orientation}. The incident direction is set as $-x$ direction,
while the impact parameter vector is set to $+y$ direction, assigning the $x$-$y$ plane
as the reaction plane. The three initial configurations considered are: \textit{i}) the symmetry
axis of $^{154}$Sm is set parallel to the collision axis ($x$ axis), \textit{ii}) the symmetry axis
is set parallel to the impact parameter vector ($y$ axis), and \textit{iii}) the symmetry axis is
set perpendicular to the reaction plane. Henceforth, we refer to those configurations as
$x$-, $y$-, and $z$-direction cases, respectively, throughout the paper.

%&&&&&&&&&&&&&&&&&&&&&&&&&&&&&&&&&&&&&&&&&&&&&&&&&&
%&&&&&&&&&&&&&&&&&&&&&&&&&&&&&&&&&&&&&&&&&&&&&&&&&&

\begin{figure}[t]
\begin{center}
\includegraphics[width=8.6cm]{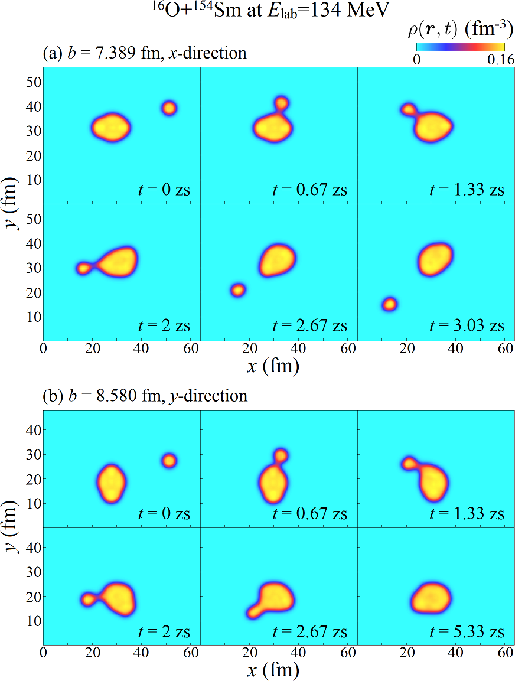}
\end{center}\vspace{-4mm}
\caption{
Snapshots of the density of the colliding nuclei in the reaction plane
obtained from the TDHF calculations for the $^{16}$O+$^{154}$Sm
reaction at $E_{\rm lab}$\,=\,134\,MeV. In panels (a), those for the
$x$-direction case with $b$\,=\,7.389\,fm (just outside the fusion threshold)
are shown; while those for the $y$-direction case with $b$\,=\,8.580\,fm
(the maximum impact parameter for fusion) are shown in panels (b).
}
\label{rho(t)}
\end{figure}

Since the incident energies examined are above the Coulomb barrier,
fusion reactions take place at small impact parameters. By repeating
TDHF calculations the maximum impact parameters for fusion, say
$b_{\rm fus}$, were identified with 0.001-fm accuracy. For the $x$-,
$y$-, and $z$-direction cases, respectively, we found: $b_{\rm fus}$\,=\,5.400,
6.090, and 5.025\,fm for $E_{\rm lab}$\,=\,85\,MeV and
$b_{\rm fus}$\,=\,7.388, 8.580, and 7.208\,fm for $E_{\rm lab}$\,134\,~MeV.
For smaller incident energies, the system requires smaller impact parameters
to fuse. Also, one may notice that the fusion reaction takes place at a larger
impact parameter for the $y$-direction case, as compared to the other two,
which can be simply understood as a geometric effect (cf. Figs.~\ref{orientation}
and \ref{rho(t)}).

In Fig.~\ref{rho(t)}, we present snapshots of the density of the colliding nuclei in
the reaction plane for the $^{16}$O+$^{154}$Sm reaction at $E_{\rm lab}$\,=\,134\,MeV.
We show two representative cases of $x$- and $y$-direction cases in panels (a) and (b),
respectively. The elapsed time in the simulation is indicated in zeptoseconds (1~zs\,=\,10$^{-21}$\,sec).
In panels (a), reaction dynamics are shown for the impact parameter of $b$\,=\,7.389\,fm,
which is just outside the maximum impact parameter for fusion for the $x$-direction case,
$b_{\rm f}$\,=\,7.388\,fm. In this case, $^{16}$O collides with the side of $^{154}$Sm
($t$\,=\,0.67\,zs), forming a neck structure. Then, the projectile-like subsystem moves
along the edge of the target-like subsystem ($t$\,=\,0.67--2\,zs), and finally reseparates
forming a binary fragments ($t$\,=\,3.03\,zs). It is interesting to notice that the shape of
the target-like fragment is distorted through the dynamic interaction during the collision.
On the other hand, in panels (b), we show reaction dynamics for the $y$-direction case
at the maximum impact parameter for fusion, $b_{\rm f}$\,=\,8.580\,fm. In this case,
$^{16}$O collides with the tip of $^{154}$Sm ($t$\,=\,0.67\,zs). As time evolves
($t$\,=\,1.33--2.67\,zs), the system develops a neck, and eventually gets fused, forming
a mononuclear shape ($t$\,=\,5.33\,zs). Figure~\ref{rho(t)} explains why in the $y$-direction
case the two nuclei interact at relatively large impact parameters, while for $x$- and
$z$-direction cases nucleon transfer takes place at smaller impact parameters.

%&&&&&&&&&&&&&&&&&&&&&&&&&&&&&&&&&&&&&&&&&&&&&&&&&&
%&&&&&&&&&&&&&&&&&&&&&&&&&&&&&&&&&&&&&&&&&&&&&&&&&&

%&&&&&&&&&&&&&&&&&&&&&&&&&&&&&&&&&&&&&&&&&&&&&&&&&&
\begin{figure*}[tb]
\begin{center}
\includegraphics[width=0.95\textwidth]{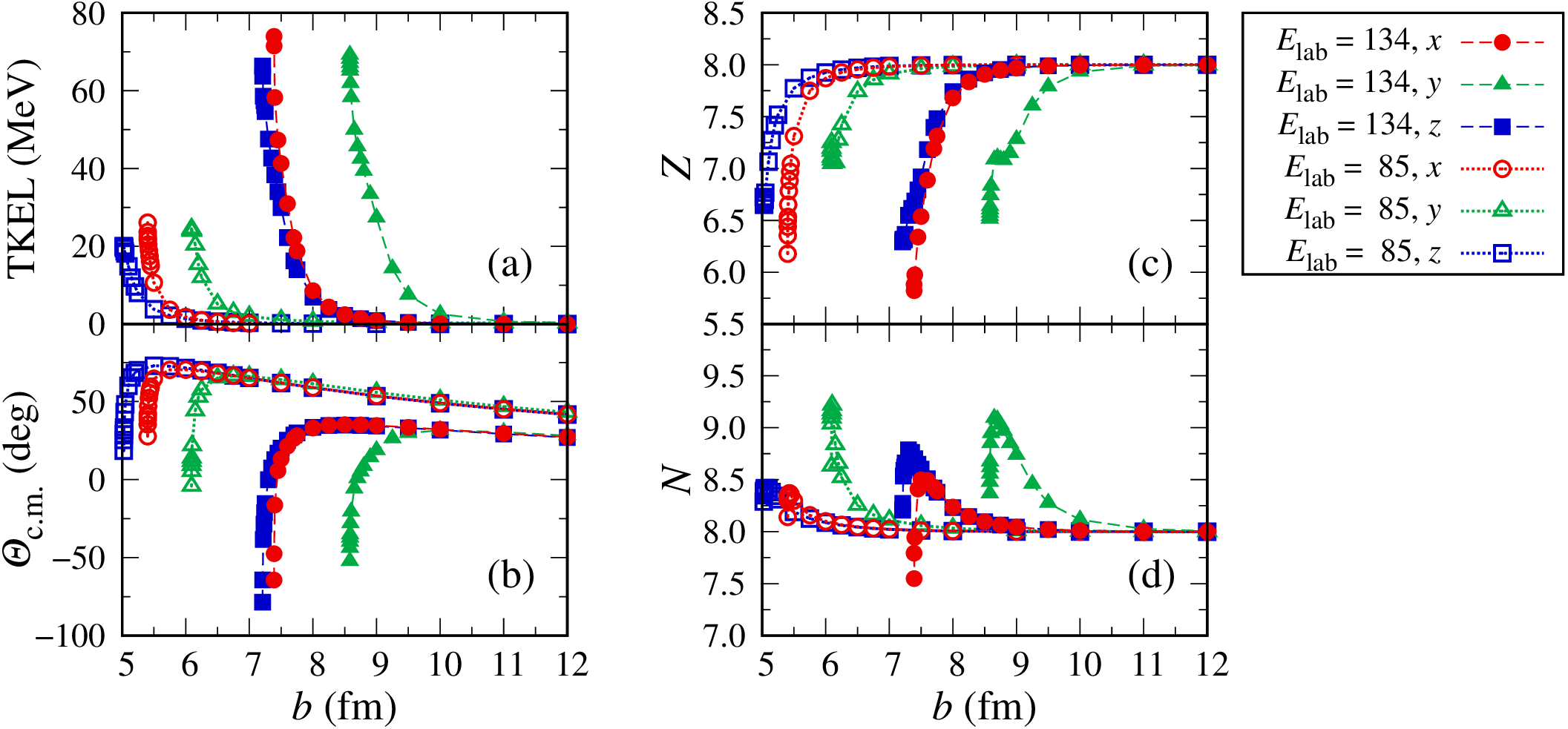}
\caption{
Results of TDHF calculations for the $^{16}$O+$^{154}$Sm reaction at $E_{\rm lab}$\,=\,85
and 134\,MeV. Results for $E_{\rm lab}$\,=\,85\,MeV are shown by open
symbols connected with dotted lines, while those for $E_{\rm lab}$\,=\,134\,MeV
are shown by solid symbols connected with dashed lines. In panels
\ref{Fig:overview}(a--d), total kinetic energy loss (TKEL), deflection
function, the numbers of protons and neutrons in PLF are shown, respectively,
as functions of the impact parameter $b$.
}
\label{Fig:overview}
\end{center}
\end{figure*}
%&&&&&&&&&&&&&&&&&&&&&&&&&&&&&&&&&&&&&&&&&&&&&&&&&

Let us now look into global features of the reaction dynamics.
In Fig.~\ref{Fig:overview}, we show TKEL (a), the deflection function (b), and the
average numbers of protons (c) and neutrons (d) in PLFs, as functions of the impact
parameter. Note that we plot both results for $E_{\rm lab}$\,=\,85\,(134)\,MeV,
which are represented by open (solid) symbols connected with dotted
(dashed) lines. Data points with circles, triangles, and squares correspond
to the $x$-, $y$-, and $z$-direction cases, respectively.

From Fig.~\ref{Fig:overview}(a), TKEL is observed to be very small when
the impact parameter is large ($b$\,$\gtrsim$\,7\,fm for $E_{\rm lab}$\,=\,85\,MeV
and $b$\,$\gtrsim$\,10\,fm for $E_{\rm lab}$\,=\,134\,MeV), corresponding to (quasi)elastic
scattering. As the impact parameter decreases, TKEL increases rapidly and
reaches maximum values as large as, e.g., 70~MeV for $E_{\rm lab}$\,=\,134\,MeV,
meaning that a large amount of kinetic energy is dissipated into internal excitations
after multinucleon exchanges. Around this impact parameter, the deflection
function decreases noticeably as shown in Fig.~\ref{Fig:overview}(b). Large
negative deflection angles correspond to large scattering angles, where the
trajectory of colliding nuclei is distorted by the nuclear attractive interaction.
The results shown in Figs.~\ref{Fig:overview}(a) and \ref{Fig:overview}(b)
exhibit noticeable orientation dependence. Namely, TKEL (deflection function)
in the $y$-direction case increases (decreases) at larger impact parameters
as compared to the other cases. The latter is solely because of the geometry,
as depicted in Fig.~\ref{rho(t)}. We note that although the results look energy
dependent as a function of the impact parameters, they show similar behavior
when one plots as a function of the distance of the closest approach (not shown
here). Thus, global feature of the reaction dynamics is almost the same for the two
energies examined. Apart from the apparent geometric effect, the orientation
of $^{154}$Sm does not alter the mean values of TKEL and $\Theta_{\rm c.m.}$.
It is to mention here that recent comprehensive measurements manifest good
agreement between experimental data and mean TKEL and scattering angles
in TDHF for deep-inelastic collisions of $^{58}$Ni+$^{58}$Ni \cite{Williams(2018)}.

Next, let us look at the average numbers of protons and neutrons in PLFs
shown in Figs.~\ref{Fig:overview}(c) and \ref{Fig:overview}(d), respectively.
As mentioned in Introduction, possible orientation effects on transfer dynamics were
observed in our preceding studies for $^{18}$O+$^{206}$Pb \cite{Bidyut(2015)}
and $^{16}$O+$^{27}$Al \cite{Bidyut(2018)}. We were thus particularly interested
in understanding the effect of the large target deformation on transfer dynamics.
As shown in Figs.~\ref{Fig:overview}(c) and \ref{Fig:overview}(d), the average numbers
of transferred nucleons are vanishingly small when the impact parameter is large
($b$\,$\gtrsim$\,7\,fm for $E_{\rm lab}$\,=\,85\,MeV and $b$\,$\gtrsim$\,10\,fm for
$E_{\rm lab}$\,=\,134\,MeV), in a similar manner as was observed for TKEL and the
deflection angle shown in Figs.~\ref{Fig:overview}(a) and \ref{Fig:overview}(b).
As the impact parameter decreases, nucleons are transferred towards the directions
of the charge equilibrium of the system, reducing the $N/Z$ asymmetry between
projectile and target nuclei. Since the initial $N/Z$ ratios are 1 for $^{16}$O and
about 1.48 for $^{154}$Sm, neutrons tend to transfer from $^{154}$Sm to
$^{16}$O, while the trend is opposite for protons. Again, we find the geometric
effect also on the transfer dynamics, where nucleon transfer occurs at relatively
larger impact parameters in the $y$-direction case as compared to the other cases.

At small impact parameters close to the fusion threshold, an abrupt change of
transfer dynamics is observed, especially for $E_{\rm lab}$\,=\,134\,MeV. Namely,
the average number of neutrons in PLFs decreases sharply [Fig.~\ref{Fig:overview}(d)],
which also accompanies a sudden decrease of the number of protons in a correlated manner
[Fig.~\ref{Fig:overview}(c)]. This kind of transfer dynamics has been routinely observed
in TDHF for various systems \cite{KS_KY_MNT,KS_KY_Ni-U,KS_U-Sn,Bidyut(2015),Bidyut(2018)},
which can be interpreted as nucleon transfer associated with neck breaking dynamics. When
two nuclei collide a neck structure is developed, during which the system evolves quickly towards
the charge equilibrium through fast exchange of neutrons and protons. When the neck dissociates
a correlated transfer of neutrons and protons occurs as a result of transfer of charge equilibrated
matter inside the neck region. For this system, a strong tendency towards fusion gives rise to the
correlated transfer of nucleons from $^{16}$O to $^{154}$Sm. In the lower incident energy case
($E_{\rm lab}$\,=\,85\,MeV) the abrupt change is hardly seen (symptom is visible though), because
the system could not overcome a strong tendency towards fusion after the neck formation, whereas
centrifugal force enabled the system to reseparate for $E_{\rm lab}$\,=\,134\,MeV.

%&&&&&&&&&&&&&&&&&&&&&&&&&&&&&&&&&&&&&&&&&&&&&&&&&&
\begin{figure*}[t]
\begin{center}
\includegraphics[width=0.95\textwidth]{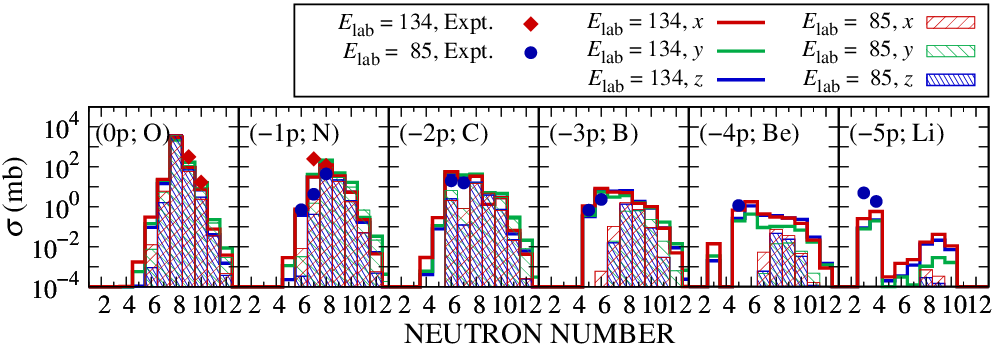}
\caption{
The $Q$-value- and angle-integrated isotope production cross sections
for various proton-transfer channels in the $^{16}$O+$^{154}$Sm reaction
at $E_{\rm lab}$\,=\,85 and 134\,MeV. The change in the number of protons
compared with that of the projectile ($Z$\,=\,8) is indicated as ($-x$p;~X),
where X stands for the corresponding element. Filled circles (diamonds) show
the experimental data for $E_{\rm lab}$\,=\,85\,(134)\,MeV. The results of
TDHF+GEMINI calculations for $E_{\rm lab}$\,=\,85\,(134)\,MeV are shown by
shaded (open) histograms, where red, green, and blue colors are used for
the $x$-, $y$-, and $z$-direction cases, respectively.
}
\label{Fig:sigmatot}
\end{center}
\end{figure*}
%&&&&&&&&&&&&&&&&&&&&&&&&&&&&&&&&&&&&&&&&&&&&&&&&&

Concerning orientation dependence of transfer dynamics, other than the geometric effect,
we find that neutron transfer from $^{154}$Sm to $^{16}$O is somewhat enhanced for the
$y$-direction case [Fig.~\ref{Fig:overview}(d)], while proton transfer from $^{16}$O to
$^{154}$Sm is enhanced for the $x$- and $z$-direction cases [Fig.~\ref{Fig:overview}(c)].
We will come back to this point in the end of this section, along with an additional theoretical
analysis of the $^{24}$O+$^{154}$Sm reaction.

Let us now compare measurements and theoretical calculations for
total ($Q$-value- and angle-integrated) isotope production cross sections.
In TDHF, transfer probabilities can be extracted from a TDHF wave function
after collision with the use of the particle-number projection method as
described in, e.g., Refs.~\cite{PNP,KS_KY_MNT}. Transfer cross sections
are then calculated by integrating the transfer probabilities over the impact
parameter. Those cross sections correspond to production of reaction
products just after reseparation before secondary deexcitation processes.
As the primary reaction products populated through MNT reactions could
be highly excited, secondary deexcitation processes via light particle emissions
may contribute significantly in the final isotope production cross sections.
Such calculations are performed in a method called TDHF+GEMINI \cite{KS_GEMINI},
which combines TDHF calculations with a statistical model, \texttt{GEMINI++}
\cite{GEMINI++,Charity(2010),Mancusi(2010)}. Details of the calculations
are as given in Ref.~\cite{KS_GEMINI}, while the experimental masses
have been updated to AME2020 \cite{AME2020-1,AME2020-2}. Because
of the large mass asymmetry of the system under study, the assumption
of thermal equilibrium, that is, distributing the total excitation energy
proportional to the fragment masses, substantially weakens the effect
of secondary evaporation from PLFs. As the experimental data indicate
large effects of secondary particle emissions, we distributed the excitation
energy equally to the fragments, i.e.
$E_{N,Z}^*(b) = \frac{1}{2}[E_{\rm c.m.}-E_{\rm kin}^\infty(b)+Q_{\rm gg}(N,Z)]$,
which results in larger evaporation effects. Here, $E_{\rm kin}^\infty(b)$ denotes
the asymptotic value of total kinetic energy (TKE) of outgoing fragments for
average products in TDHF at an impact parameter $b$ and $Q_{\rm gg}$
is the ground-to-ground $Q$ value for the exit channel involving a nucleus
with $N$ neutrons and $Z$ protons.

In Fig.~\ref{Fig:sigmatot}, we compare the measured integrated cross sections
with the results by TDHF+GEMINI. Filled diamonds (filled circles) represent the measured cross sections
for $E_{\rm lab}$\,=\,134\,(85)\,MeV. The results of TDHF+GEMINI for $E_{\rm lab}$\,=\,134\,MeV
are shown by open histograms, while those for $E_{\rm lab}$\,=\,85\,MeV are shown
by shaded histograms. Red, green, and blue colors correspond to the results associated
with the $x$-, $y$-, and $z$-direction cases, respectively. The cross sections are
classified according to the number of transferred protons as indicated by ($-x$p;\,X),
where X stands for the corresponding element, and are plotted as a function of
the neutron number of the reaction product.

In the case of $E_{\rm lab}$\,=\,134\,MeV, the experimental data are limited for
($0$p) and ($-1$p) channels with transfer of few neutrons (cf.~Figs.~\ref{AngDistSm}
and \ref{SigmaInt154Sm}). The agreement between theoretical predictions and
measurements is reasonable, although the data are limited and TDHF provides
somewhat smaller cross sections, especially for the ($-1$p,\,$-1$n) channel.

In the case of $E_{\rm lab}$\,=\,85\,MeV, cross sections were measured for
more abundant proton-transfer channels, as compared to the $E_{\rm lab}$\,=\,134\,MeV
case. For transfer of a few nucleons, i.e., ($-1$p,\,$0$n) and ($-1$p,\,$-1$n),
we again find reasonable agreement between the theoretical and experimental
results. For other measured cross sections, however, TDHF substantially underestimates
the experimental data, even after the inclusion of secondary deexcitation effects with
\texttt{GEMINI++} (compare filled circles and shaded histograms). On the other hand,
we find that the cross sections for secondary products associated with $E_{\rm lab}$\,=\,134\,MeV
(open histograms) follow nicely the trends observed in the experimental data for $E_{\rm lab}$\,=\,85\,MeV.
Especially, production of $^{10,11}$B, $^9$Be, and $^{6,7}$Li can only be explained
by light-charged-particle emissions from excited PLFs. It indicates that the effects of
secondary deexcitation processes, i.e., excitation energies, may be underestimated
for $E_{\rm lab}$\,=\,85\,MeV in the present TDHF+GEMINI analysis.
Indeed, the TKEL value for mean binary products in $E_{\rm lab}$\,=\,85\,MeV
was at most 30\,MeV in TDHF [cf.\ Fig.~\ref{Fig:overview}(a)], whereas experimental
TKEL distributions for $^{10,11}$B, $^9$Be, and $^{6,7}$Li show substantial deep-inelastic
components around TKEL\,$\approx$\,40--60\,MeV, which indicates importance
of secondary light-particle emissions.

The underestimation of excitation energy may be originated from the mean-field nature
of the TDHF approach. To evaluate excitation energy of reaction products, mean total
excitation energy was distributed equally to the binary products. On one hand, it has been
well established that TDHF can provide quantitative description of the most probable reaction
outcomes, such as average total excitation energy. On the other hand, in reality, the excitation
energy has certain distribution, meaning that part of fragments can have higher excitation
energies than the mean value, for which larger evaporation effects are expected. To improve
the description, one may employ a theoretical framework that incorporates beyond-TDHF
fluctuations and correlations, e.g., time-dependent random phase approximation (TDRPA)
\cite{Simenel(review)} which is based on an extended variational principle of Balian and V\'en\'eroni
\cite{BV(1981)} or stochastic mean-field (SMF) approach \cite{Ayik(2008),Lacroix(2014),KS_AS_Ni-Pb}.
A work on kinetic energy distributions within the SMF approach is in progress \cite{AS_KS_Xe-Pb}.

Overall, it turns out that the orientation effects in MNT processes in the
$^{16}$O+$^{154}$Sm system is rather small, which is hard to disentangle
from the integrated production cross sections, as shown in Fig.~\ref{Fig:sigmatot}.
We point out that the incident energies examined here are both above the Coulomb
barrier for all orientations. By decreasing the incident energy, for instance, below
the Coulomb barrier for collisions from the equatorial side of $^{154}$Sm, but
above the barrier for collisions from the tip of $^{154}$Sm, the main contribution
to the MNT processes would come from the latter, deformation aligned configuration.
By carefully choosing the incident energy, one could at least investigate such a
selective orientation effect on transfer processes. Decreasing the incident energy
further to the subbarrier regime for all orientations, transfer processes are dominated
by quantum tunneling of nucleonic wave functions. In such a case, transfer probabilities
would be more sensitive to the nature of single-particle orbitals, such as spatial distribution
and angular momentum. We leave this further exploration of possible orientation
effects on MNT processes in subbarrier regime as a future work.

Finally, to serve additional information on the reaction mechanism, we present
a purely theoretical investigation of isotope dependence of transfer dynamics, taking
the $^{24}$O+$^{154}$Sm reaction as an example. For the $^{16}$O+$^{154}$Sm system,
neutron pickup and proton stripping are favored because of the initial $N/Z$ asymmetry
of the system. By replacing $^{16}$O with $^{24}$O, where the latter nucleus has a large
$N/Z$ ratio, $N/Z$\,=\,2, we can revert the favored directions of neutron and proton
transfers as compared to the $^{16}$O+$^{154}$Sm case. This allows us to further examine
effects of single-particle structures, since properties of `donor' and `acceptor' orbitals
are then different. Also, it serves useful information for future experiments with
neutron-rich radioactive-ion beam. For the $^{24}$O+$^{154}$Sm system, the fusion threshold
impact parameters are found to be: $b_{\rm fus}$\,=\,5.807, 6.583, and 5.530\,fm for
$E_{\rm c.m.}$\,=\,77\,MeV and $b_{\rm fus}$\,=\,7.744, 8.961, and 7.571\,fm for $E_{\rm c.m.}$\,=\,121\,MeV.
These impact parameters are slightly larger than those for the $^{16}$O+$^{154}$Sm system
because of the excess neutrons in $^{24}$O.

\begin{figure}[tb]
\begin{center}
\includegraphics[width=\columnwidth]{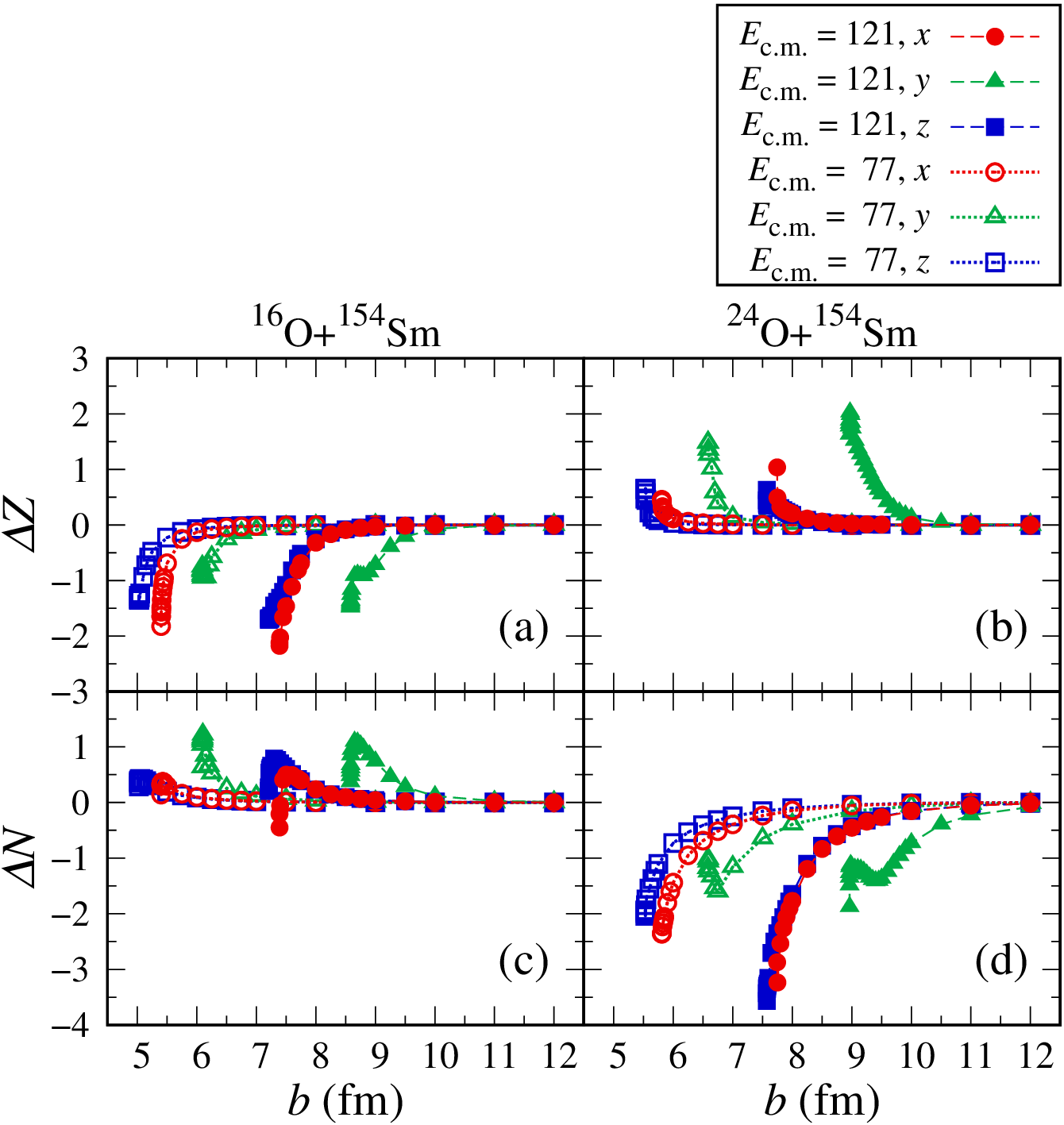}
\end{center}\vspace{-3mm}
\caption{
Results of TDHF calculations for the $^{16,24}$O+$^{154}$Sm reactions
at $E_{\rm c.m.}$\,=\,77 and 121\,MeV. The average numbers of transferred
protons (neutrons) are shown in upper (lower) panels as functions of the
impact parameter $b$. Plus (minus) sign corresponds to addition to (removal
from) the projectile nucleus. In the left column [panels (a) and (c)], results
for the $^{16}$O+$^{154}$Sm reaction are shown [the same data shown
in Figs.~\ref{Fig:overview}(c) and \ref{Fig:overview}(d)], while in the right
column [panels (b) and (d)] shows those for the $^{24}$O+$^{154}$Sm
reaction. Results for $E_{\rm c.m.}$\,=\,77\,MeV are shown by open symbols
connected with dotted lines, while those for $E_{\rm c.m.}$\,=\,121\,MeV are
shown by solid symbols connected with dashed lines.
}
\label{Fig:24O+154Sm}
\end{figure}

In Figs.~\ref{Fig:24O+154Sm}(b) and \ref{Fig:24O+154Sm}(d), we show the average
number of transfered protons and neutrons, respectively, with respect to the projectile as
a function of the impact parameter for the $^{24}$O+$^{154}$Sm reaction at two incident
energies. The incident energies were chosen to give the same center-of-mass energies as
the $^{16}$O+$^{154}$Sm reaction at $E_{\rm lab}$\,=\,85 and 134\,MeV.
In Figs.~\ref{Fig:24O+154Sm}(a) and \ref{Fig:24O+154Sm}(c), we also present the
same plot for the $^{16}$O+$^{154}$Sm system [the same data shown in
Figs.~\ref{Fig:overview}(c) and \ref{Fig:overview}(d)] for comparison. The lower energy
case ($E_{\rm c.m.}$\,=\,77\,MeV) is shown by open symbols connected with dotted lines,
while the higher energy case ($E_{\rm c.m.}$\,=\,121\,MeV) is shown by solid symbols connected
with dashed lines. Colors and symbols are the same as Fig.~\ref{Fig:overview}.

%&&&&&&&&&&&&&&&&&&&&&&&&&&&&&&&&&&&&&&&&&&&&&&&&&&
%&&&&&&&&&&&&&&&&&&&&&&&&&&&&&&&&&&&&&&&&&&&&&&&&&&

From the figure, one can clearly see that the directions of neutron and proton transfers are
indeed reverted by replacing $^{16}$O with $^{24}$O, as expected (compare the left and
right panels). From Fig.~\ref{Fig:24O+154Sm}(b), we find that proton transfer from
$^{154}$Sm to $^{24}$O is somewhat enhanced for the $y$-direction case (green triangles),
where $^{24}$O collides with a point close to the tip of $^{154}$Sm [cf.\ Fig.~\ref{rho(t)}(b)],
which is not observed for proton transfer in the $^{16}$O+$^{154}$Sm reaction [Fig.~\ref{Fig:24O+154Sm}(a)].
From a careful look at neutron transfer from $^{154}$Sm to $^{16}$O shown in Fig.~\ref{Fig:24O+154Sm}(c)
a similar enhancement in the $y$-direction case is seen. On the other hand, from Fig.~\ref{Fig:24O+154Sm}(d),
we find that neutron transfer from $^{24}$O to $^{154}$Sm looks larger for the $x$- and $z$-direction
cases, where $^{24}$O collides close to the equatorial side of $^{154}$Sm [cf. Fig.~\ref{rho(t)}(a)].
Similar tendency is also observed in proton transfer from $^{16}$O to $^{154}$Sm in the $^{16}$O+$^{154}$Sm
reaction [Fig.~\ref{Fig:24O+154Sm}(a)]. Combining these observations for the $^{16}$O+$^{154}$Sm
and $^{24}$O+$^{154}$Sm systems, we may conclude that: \textit{i}) addition of nucleons
to the prolately deformed $^{154}$Sm may be enhanced when the projectile collides with
the equatorial side of $^{154}$Sm, and \textit{ii}) removal of nucleons from $^{154}$Sm
may be favored when the projectile collides with the tip of $^{154}$Sm. Certainly, transfer
dynamics reflect structural properties of `donor' and `accepter' nuclei. By carefully choosing
colliding nuclei and analyzing transfer processes, one may be able to investigate effects of
single-particle orbitals as well as deformation on multinucleon transfer processes in low-energy
heavy-ion reactions.

\section{SUMMARY}\label{sec:summary}

There has been a revival of interest on multinucleon transfer reactions in recent years
not only as a way to investigate nucleon-nucleon correlations and single-particle properties,
but also as a possible means to produce unknown neutron-rich nuclei. Since the majority
of atomic nuclei manifest static deformation in their ground state, it is natural to investigate
the role of mutual orientations of deformed nuclei in the multinucleon transfer mechanism.

To develop understanding of the reaction mechanism and to shed some light on the effect
of nuclear orientation on multinucleon transfer processes, we have conducted experiments on
the $^{16}$O+$^{154}$Sm reaction at $E_{\rm lab}$\,=\,85\,MeV (near the Coulomb barrier)
and 134\,MeV (substantially above the barrier), where the target nucleus, $^{154}$Sm, is
a well-deformed nucleus. Angular distributions for elastic scattering and for various transfer
channels as well as energy (TKEL) spectra have been measured, and $Q$-value- and angle-integrated
isotope production cross sections have been obtained.

To understand the experimental data, the $^{16}$O+$^{154}$Sm reaction has been
analyzed based on the microscopic framework of the time-dependent Hartree-Fock (TDHF)
theory. Production cross sections for various transfer channels have been obtained employing
the particle-number projection method \cite{PNP}. With the use of a statistical compound-nucleus
deexcitation model, \texttt{GEMINI++} \cite{GEMINI++,Charity(2010),Mancusi(2010)},
effects of secondary particle evaporation processes have been taken into account. From
the comparison between the theoretical and the experimental results for production
cross sections, we have found reasonable agreement for channels accompanying transfer
of a few nucleons. For the $E_{\rm lab}$\,=\,85\,MeV case, however, TDHF+GEMINI
underestimates production cross sections for many-nucleon transfer channels. On the other
hand, we have found that cross sections for the higher energy case ($E_{\rm lab}$\,=\,134\,MeV)
nicely capture the trends observed for the $E_{\rm lab}$\,=\,85\,MeV case. The latter observation
indicates possible underestimation of excitation energy of reaction products in the present
TDHF+GEMINI analysis. It would be possible to improve the description by using, e.g.,
the stochastic mean-field (SMF) approach \cite{Ayik(2008),Lacroix(2014),KS_AS_Ni-Pb,
AS_KS_Xe-Pb}. Although the present analysis certainly offers additional information on
the multinucleon transfer mechanism, however, it turned out that the orientation effect
is rather weak, hard to disentangle from the integrated cross sections, at least for the
$^{16}$O+$^{154}$Sm reaction at the two incident energies examined.

Although we could not see clear orientation effects in the integrated cross sections, some
symptom of orientation dependence of transfer dynamics has been observed in TDHF calculations
for the $^{16,24}$O+$^{154}$Sm reactions, where the $^{24}$O+$^{154}$Sm system was
additionally analyzed to explore isotope as well as orientation dependence of transfer mechanism within
the TDHF approach. As is naively expected, there is a simple geometric effect on reaction dynamics,
which makes reactions take place at larger impact parameters when the deformation axis of $^{154}$Sm
is set aligned to the impact parameter vector. Apart from the geometric effect, we have found that:
\textit{i}) addition of nucleons to the prolately deformed $^{154}$Sm (or tendency towards fusion)
may be enhanced when the projectile collides with the equatorial side of $^{154}$Sm, \textit{ii})
removal of nucleons from $^{154}$Sm may be favored when the projectile collides with the tip of
$^{154}$Sm. Although we could not disentangle such orientation effects from the present experimental
data, as a future work one may explore the sub-barrier energy regime where data would be sensitive
 to single-particle properties.

\begin{acknowledgements}
Support from the staff of BARC-TIFR Pelletron-Linac facility, Mumbai,
is highly appreciated for their excellent support during the beam time.
We are thankful to Drs.~Rahul Tripaty and T.N.Nag, Radio Chemistry Division, BARC,
Mumbai, for providing aluminium target and preparing samarium target for us.
We thank Drs.\ A.K.~Mohanty, P.P.~Singh, T.~Sinha, V.~Jha, A.~Kundu, D.~Chattopadhyay, Sonika, and A.~Parmar
for their help during the beam time and interest during the initial stages of this work.
One of the authors (BJR) is thankful to a project trainee Nitika for her involvement
in some parts of the data analysis as a part of her internship work. We also thank
Harun Al Rashid for his interest and contribution to this work.
Theoretical calculations were performed using computational resources of
the HPCI system (Oakforest-PACS) provided by Joint Center for Advanced
High Performance Computing (JCAHPC) through the HPCI System Project
(Project ID: hp210023). This work also used (in part) computational resources
of the Yukawa-21 System at Yukawa Institute for Theoretical Physics (YITP),
Kyoto University. One of the authors (KS) was supported by the Japan
Society for the Promotion of Science (JSPS) KAKENHI, Grant-in-Aid for
Early-Career Scientists, Grant Number: 19K14704.
\end{acknowledgements}

\end{document}